\documentclass[onecolumn,showpacs,preprintnumbers,amsmath,amssymb,superscriptaddress]{revtex4}

\usepackage{epsf}
\usepackage{graphicx}
\usepackage{sidecap}

 \usepackage{array}
 \usepackage{amsmath}
 \usepackage{amssymb}
 \usepackage{dcolumn}
 \usepackage{epstopdf}
 \usepackage{bm}% bold math
 \usepackage{amsmath}

\usepackage{color}
 
\def \beq {\begin{equation}}
\def \eeq {\end{equation}}
\begin{document}
 
\title{{Observation of gapped state in rare-earth monopnictide HoSb}}
 
\author{M.~Mofazzel~Hosen}\affiliation {Department of Physics, University of Central Florida, Orlando, Florida 32816, USA}
\author{Gyanendra~Dhakal}\affiliation {Department of Physics, University of Central Florida, Orlando, Florida 32816, USA}
\author{Baokai Wang}\affiliation{Department of Physics, Northeastern University, Boston, Massachusetts 02115, USA}
\author{Narayan Poudel}\affiliation{Idaho National Laboratory, Idaho Falls, Idaho 83415, USA}
\author{Bahadur Singh}\affiliation{Department of Physics, Northeastern University, Boston, Massachusetts 02115, USA}
\author{Klauss~Dimitri}\affiliation {Department of Physics, University of Central Florida, Orlando, Florida 32816, USA}
\author{Firoza~Kabir}\affiliation {Department of Physics, University of Central Florida, Orlando, Florida 32816, USA} 
\author{Christopher~Sims}\affiliation {Department of Physics, University of Central Florida, Orlando, Florida 32816, USA}

\author{Sabin~Regmi}\affiliation {Department of Physics, University of Central Florida, Orlando, Florida 32816, USA}
\author{William~Neff}\affiliation {Department of Physics, University of Central Florida, Orlando, Florida 32816, USA}
\author{Anan Bari Sarkar}
\affiliation{Department of Physics, Indian Institute of Technology, Kanpur 208016, India}
\author {Amit Agarwal}
\affiliation{Department of Physics, Indian Institute of Technology, Kanpur 208016, India}
\author{Daniel Murray}\affiliation{Idaho National Laboratory, Idaho Falls, Idaho 83415, USA}
\author{Franziska Weickert}\affiliation{National High Magnetic Field Laboratory, Los Alamos, New Mexico, 87545, USA}
%\author{Tomasz~Durakiewicz}
%\affiliation {Condensed Matter and Magnet Science Group, Los Alamos National Laboratory, Los Alamos, NM 87545, USA}
%\affiliation {Institute of Physics, Maria Curie - Sk{\l}odowska University, 20-031 Lublin, Poland}
\author{Krzysztof Gofryk}\affiliation{Idaho National Laboratory, Idaho Falls, Idaho 83415, USA}
\author{Orest Pavlosiuk} \affiliation{Institute of Low Temperature and Structure Research, Polish Academy of Sciences, 50-950  Wroc{\l}aw, Poland}
\author{Piotr Wi{\'s}niewski} \affiliation{Institute of Low Temperature and Structure Research, Polish Academy of Sciences, 50-950  Wroc{\l}aw, Poland}
\author{Dariusz~Kaczorowski}\affiliation{Institute of Low Temperature and Structure Research, Polish Academy of Sciences, 50-950  Wroc{\l}aw, Poland}
\author{Arun~Bansil}
\affiliation {Department of Physics, Northeastern University, Boston, Massachusetts 02115, USA}
\author{Madhab~Neupane*}
\affiliation {Department of Physics, University of Central Florida, Orlando, Florida 32816, USA}
 
\date{\today}
\pacs{}
 
\begin{abstract}
 
{
{
The rare-earth monopnictide family is attracting an intense current interest driven by its unusual extreme magnetoresistance (XMR) property and the potential presence of topologically non-trivial surface states. The experimental observation of non-trivial surface states in this family of materials are not ubiquitous. Here, using high-resolution angle-resolved photoemission spectroscopy (ARPES), magnetotransport, and parallel first-principles modeling, we examine the nature of electronic states in HoSb. Although we find the presence of bulk band gaps at the $\Gamma$ and $X$-symmetry points of the Brillouin zone (BZ), we do not find these gaps to exhibit band inversion so that HoSb does not host a Dirac semimetal state. Our magnetotransport measurements indicate that HoSb can be characterized as a correlated nearly-complete electron-hole-compensated semimetal. Our analysis reveals that the nearly perfect electron-hole compensation could drive the appearance of non-saturating XMR effect in HoSb.
}}
\end{abstract}
\maketitle

\noindent
\textbf{Introduction}\\
Topological insulators (TIs) with novel properties such as extreme magnetoresistance and high carrier mobility are currently attracting intense interests in condensed matter and material science communities. A TI supports the presence of gapless topological surface states (TSSs) with Dirac-cone-like energy dispersions within an inverted gap between the bulk conduction and valence bands \cite{Hasan, Xia, Hasan_review_2,RMP}. The discovery of TIs has spurred the exploration of gapless nontrivial states beyond the insulators to include a great variety of topological semimetals such as the Dirac-, Weyl-, and nodal-line/loop semimetals in which the bulk bandgap opening is prohibited by crystalline symmetries \cite{Dai, Neupane_2, Suyang_Science, Hong_Ding,Neupane_5}. These developments have opened up exciting new routes for finding exotic quantum phases and improved materials platforms for constructing low-power electronics/spintronics devices \cite{Hasan, Yoshomi}. A distinct nontrivial $\mathbb{Z}_2$ state with TSSs has been recently reported in a three-dimensional (3D) material with a vanishing global bandgap \cite{Kane, Wang} where the TSSs overlap with the bulk states. The rare-earth monopnictide (REM) family is drawing special interests as a rich playground for investigating Dirac fermionic excitations. REMs have been predicted to host topological Dirac semimetal as well as 3D TI states \cite{Zeng} and exhibit extremely large magnetoresistance (XMR) \cite{Pavlosuik, Tafti, Guo, Zhu, Shen, Yu, HoBi}. Although a good deal of work in the literature concerns the linearly-dispersing states and their role in driving XMR in the REMs, a robust conclusion in this regard remains a matter of debate. Although the conventional explanation for the XMR effects involves electron-hole compensation \cite{Tafti, Guo,Yu,HoBi, DK1, DK2}, the observation of XMR in the REM family and numerous other topological materials such as WTe$_2$ \cite{Mazhar}, Cd$_3$As$_2$ \cite{Cava1}, PtSn$_4$ \cite{Canfield}, and TaAs \cite{TaAs_transport} has ignited the possibility that XMR might have its origin in the forbidden backscattering channels of topological materials \cite{Cava1, TaAs_transport}. Notably, TSSs have been reported in LaBi, LaSb, CeSb, and CeBi by various groups \cite{Felser,Feng,Dessau,Niu,Lou,Kaminski, Ding, Oinuma, Nasser, Kondo,Guo} whereas LaAs, LuBi, YBi, YSb, and CeSb have been suggested to be topologically trivial materials \cite{Dessau, Shen, Tafti, Pavlosuik, Oinuma,Suzuki}. The presence of a Dirac semimetal state is suggested in NdSb \cite{Madhab} and DySb \cite{Zhu}. With this background in mind, further experimental and theoretical studies are needed to gain an understanding of the nature of Dirac-like states and how they are connected with XMR in the REM family.\\~\\ 
\noindent
\begin{figure*}
	\centering
	\includegraphics[width=18.35cm]{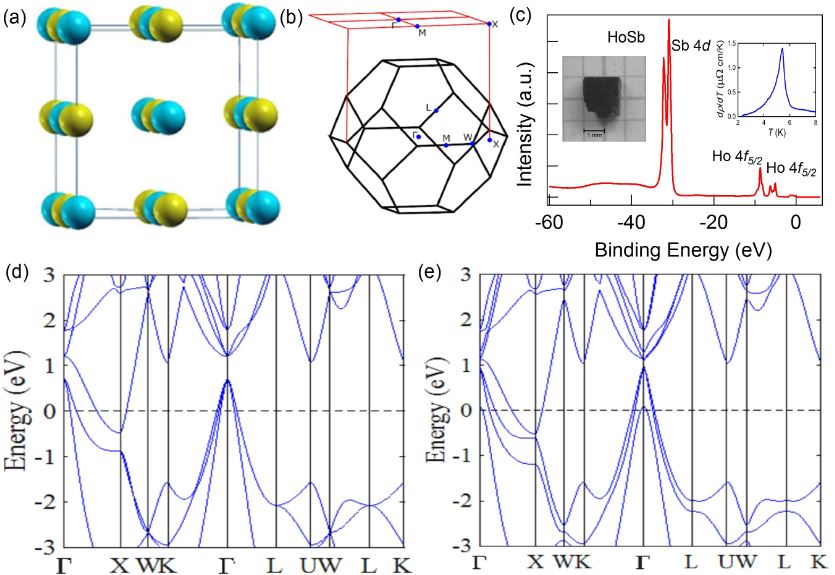}
	\caption{\textbf{Sample characterizations and electronic structure of HoSb.} (a) NaCl-type (rock salt) crystal structure of HoSb. Yellow and blue balls represent Ho and Sb atoms, respectively. (b) Primitive bulk Brillouin zone (BZ) and the projected (001) surface BZ. High-symmetry points are marked. (c) Core level spectrum of HoSb showing sharp peaks of Sb 4\textit{d} and Ho 4\textit{f}. Right inset shows the variation of the temperature derivative of resistivity with temperature, sharp peak at $\sim$5.7 K marks the magnetic transition. Left inset shows the picture of a HoSb single crystal. (d),(e) Calculated bulk band structure along the various high-symmetry directions without and with the inclusion of SOC, respectively. }
	\label{fig:CS}
\end{figure*}
\noindent
Rare-earth elements with their \textit{f}-electrons provide strongly-correlated, tunable magnetic ground states in the REM family. Moving from La to Lu in the lanthanide series, one observes a nonmagnetic to ferromagnetic transition. HoSb that supports an antiferromagnetic (AFM) ground state \cite{Vogt} is not a well-studied member of the REM family. It displays a magnetic transition from an MnO-type AFM arrangement to a HoP-type ferromagnetic arrangement under external magnetic field \cite{Mag_trans}. More recently, an unusual XMR has been reported in HoSb \cite{Xia1,Gupta}. 
Here, we report the observation of a gapped state at the ${X}$ point of the BZ in HoSb. Using angle-resolved photoemission spectroscopy (ARPES) along with first-principles calculations and magneto-transport measurements, we examine in-depth the electronic structures of this material. Our analysis reveals the presence of a highly anisotropic Dirac-like cone at the ${X}$ point. Our experimental data and theoretical results identify a small gap around 470 meV below the Fermi level. More importantly, our calculations of the $\mathbb{Z}$$_2$ invariant show that HoSb assumes a trivial topological state ($\mathbb{Z}$$_2$ = 0). We analyze our magnetotransport results to show that the presence of nearly perfect electron-hole compensation could drive the appearance of non-saturating XMR effect in HoSb.\\~\\
\noindent
\begin{figure*}
	\centering
	\includegraphics[width=18.35cm]{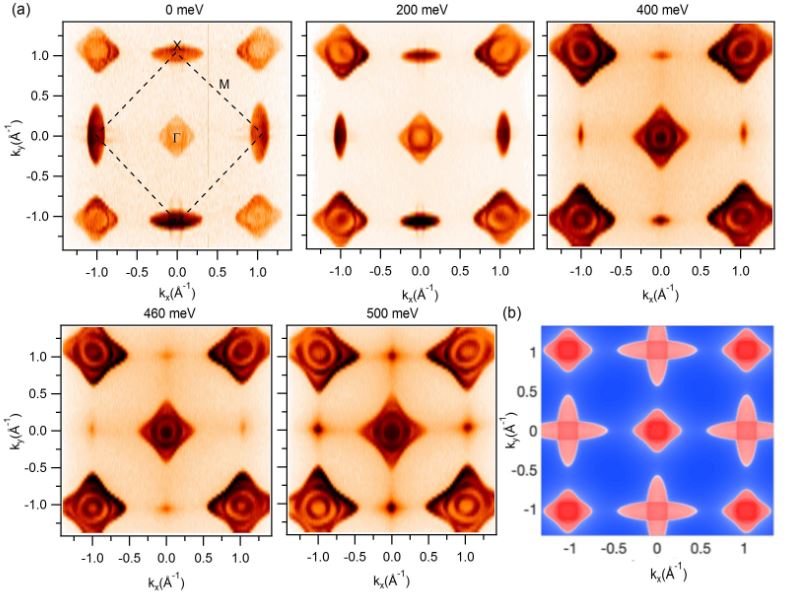}
	\caption{\textbf{Fermi surface map and constant-energy contour plots.} (a) Experimentally observed Fermi surface and constant-energy contour plots measured at a photon energy of 100 eV in HoSb. Binding energies are noted on the plots. Experiments were performed at the ALS beamline 4.0.3 at a temperature of 18 K. (b) Theoretical Fermi surface of HoSb.}
	\label{fig:CEC}
\end{figure*}
\textbf{Results}\\
\textbf{Sample characterizations and electronic structure}\\
  HoSb crystalizes in a rock-salt-type crystal structure with space group \textit{Fm-3m} like other members of the REM family (see Fig. \ref{fig:CS} (a)). Our X-ray diffraction measurements confirm the cubic crystal structure with the refined lattice parameter \textit{a} = 6.13(1) \textup{\AA}. Here, yellow and blue balls correspond to the Ho and Sb atoms, respectively (Fig. \ref{fig:CS} (a)).  Figure \ref{fig:CS} (b) shows the bulk BZ and its projection on the (001) surface. The core level spectrum is shown in Fig. \ref{fig:CS} (c). Peaks of Sb 4\textit{d} ($\sim$32 eV), Ho 4\textit{f}$_{5/2}$ ($\sim$8.6 eV) and Ho 4\textit{f}$_{3/2}$ ($\sim$5.2 eV) are clearly resolved. The observation of sharp peaks in the spectrum indicates that our HoSb samples are of good quality. The left inset of \ref{fig:CS}(c) shows a picture of our cubic crystal. In order to determine the magnetic transition temperature, we consider the temperature derivative of resistivity, $d\rho/dT$,  in the low-temperature regime (see Fig. \ref{fig:CS}(c)-right inset). One can clearly see the magnetic transition at around 5.7 K, which is in accord with previous report \cite{Mag_trans}.

We present the bulk band structure of HoSb without including spin-orbit coupling (SOC) effects and treating \textit{f}-electrons as core orbitals in Fig. \ref{fig:CS}(d). There are three hole-like bands at the $\Gamma$ point and one electron-like band at the $X$ point that crosses the Fermi level. On including SOC in the computations, salient features of this band structure picture are preserved (Fig. \ref{fig:CS}(e)), although the third hole-band at $\Gamma$ moves closer to the Fermi level so that its top cross the Fermi level. At the $X$ point, the gap between the Ho \textit{d} and Sb \textit{p} states is seen to nearly close. An inspection of the valence and conduction bands shows the presence of gaps of $\sim$ 110 meV and $\sim$ 90 meV at the $\Gamma$ and $X$ points, respectively.\\~\\
% Therefore, one would expect band inversion and a surface state within the gap to be considered this material as possible $\mathbb{Z}$$_2$-type topological semimetal.\\~\\
\begin{figure*}
	\centering
	\includegraphics[width=19.35cm]{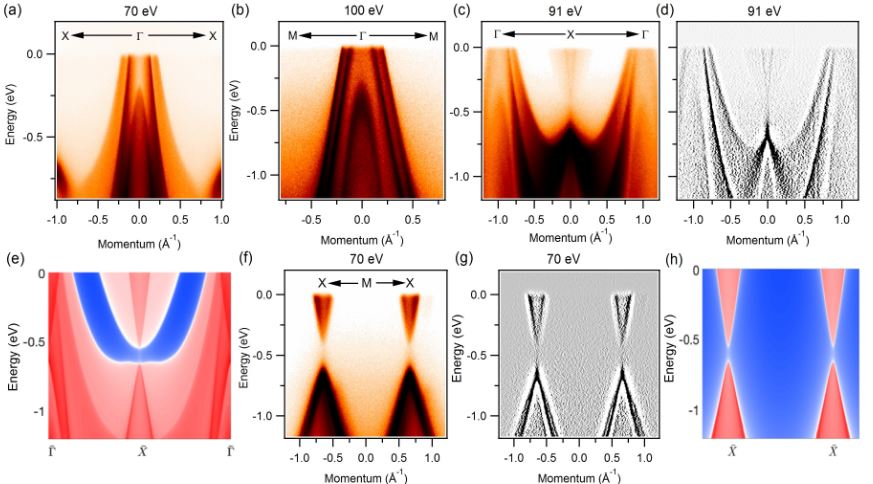}
	\caption{\textbf{Dispersion maps along the various high-symmetry directions.} (a),(b) Measured dispersion maps along the $X$-$\Gamma$-$X$ and the M-$\Gamma$-M directions. In both plots two hole-like bands can be seen to cross the Fermi level. (c),(d) Measured dispersion map and its second derivative along the $\Gamma$-$X$-$\Gamma$ symmetry lines, respectively. (e) Calculated energy dispersion along $\Gamma$-$X$-$\Gamma$. (f),(g) Experimentally measured dispersion map and its second derivative along $X$-M-$X$. Photon energies are noted on the various panels. (h) Calculated energy dispersion along the ${X}$-${X}$ direction. Experiments were performed at the ALS beamline 4.0.3 at a temperature of around 18 K. }
	\label{fig:EDC}
\end{figure*}
\textbf{Fermi surface and constant-energy contour plots}\\
We now discuss the bulk band structure and the Fermi surface using our measured constant energy contours for the (001) surface, see Fig. \ref{fig:CEC}(a). $\Gamma$ denotes the center, X the corner and M the midpoint between the two corners of the BZ. At the $\Gamma$ point, we observe a diamond-like outer Fermi pocket and a circular inner pocket. Moving towards higher binding energies, we observe a third band and that the size of the pockets increases, confirming the hole-like nature of the bands around $\Gamma$. At the $X$ point, we clearly see two concentric elliptical-shaped pockets. However, moving to higher binding energies, the elliptical-pockets evolve into point-like features around 460 meV, which indicates the electron-like nature of the bands around the $X$ point.  Importantly, our calculations show the presence of a local band gap near this binding energy at the $X$ point. Around the binding energy of 600 meV, size of the elliptical pocket increases, suggesting that the possible bulk bandgap lies around this binding energy. We will return below to present additional photon energy-dependent dispersion maps around the $\Gamma$ and $X$ points in order to ascertain the nature of the bands from our experimental measurements. Figure \ref{fig:CEC}(b) shows the calculated Fermi surface contour, which is in substantial agreement with our experimental results.\\~\\

\textbf{Trivial electronic structures of HoSb}\\
Figure \ref{fig:EDC} presents energy dispersion maps along the high-symmetry directions in the BZ. Figure \ref{fig:EDC}(a) shows the energy dispersion along the $X-\Gamma-X$ direction measured at a photon energy of 70 eV. Two hole-bands are observed to cross the Fermi level while the top of a third hole-band can be seen at around 200 meV below the Fermi level. Figure \ref{fig:EDC}(b) shows the measured dispersion map around the $\Gamma$ point along the high-symmetry direction $M-\Gamma-M$. Similar to Fig. \ref{fig:EDC}(a), three hole-bands are observed with two bands crossing the Fermi level. Importantly, two bands crossing the Fermi level along the $\Gamma-M$ direction are sharply dispersive in comparison to the $\Gamma-X$ direction, which indicates the highly anisotropic nature of these bands and also explains the distortion of the diamond-shaped pocket as we move towards higher binding energies in Fig. \ref{fig:CEC}.  Figures \ref{fig:EDC}(c) and,(d) show the dispersion map and the second derivative plots of the spectra along the $\Gamma-X-\Gamma$ direction at a photon energy of 91 eV. One can observe the gapped Dirac-like state at the X points of the BZ. The bulk band gap at the X point is better resolved in the second derivative plot (see also Supplementary Fig. 3(b)). Notably, the preceding experimental results are in substantial accord with our theoretically predicted dispersions in Fig. \ref{fig:EDC}(e).
 %This observation indicates that, HoSb can not be a topological Dirac semimetal. Furthermore, our parity calculations confirms that no band inversion takes place along the $\Gamma$-X direction. These experimental observations are in substantial agreement with our calculated (001) energy dispersion in Fig. \ref{fig:EDC}(e).  Therefore, above observations also negate HoSb as a possible $\mathbb{Z}$$_2$-type topological semimetal.
  We further confirm this by presenting photon energy dependent dispersion nature of the bands along the X-M-X direction and its second derivative in Figs. \ref{fig:EDC}(f), (g) (see also Supplementary Fig. 5), respectively. The experimentally observed bulk band gap is approximately 120 meV. Figure \ref{fig:EDC}(h) shows that the  calculated dispersion along the $X$-$X$ line is in excellent accord with the correponding experimental results. The theoretical bulk band gap at the $X$ point is about 90 meV.  There is a continuous band gap between the valence and conduction bands so that the $\mathbb{Z}$$_2$ invariant is well defined.  Using the evolution of Wannier charge centers \cite{Z2}, we find a trivial topological invariant $\mathbb{Z}$$_2$=0. In this way, we conclude that HoSb is a trivial semimetal and not a topological semimetal.\\~\\
\noindent
\begin{figure*}
	\centering
	\includegraphics[width=17.5cm]{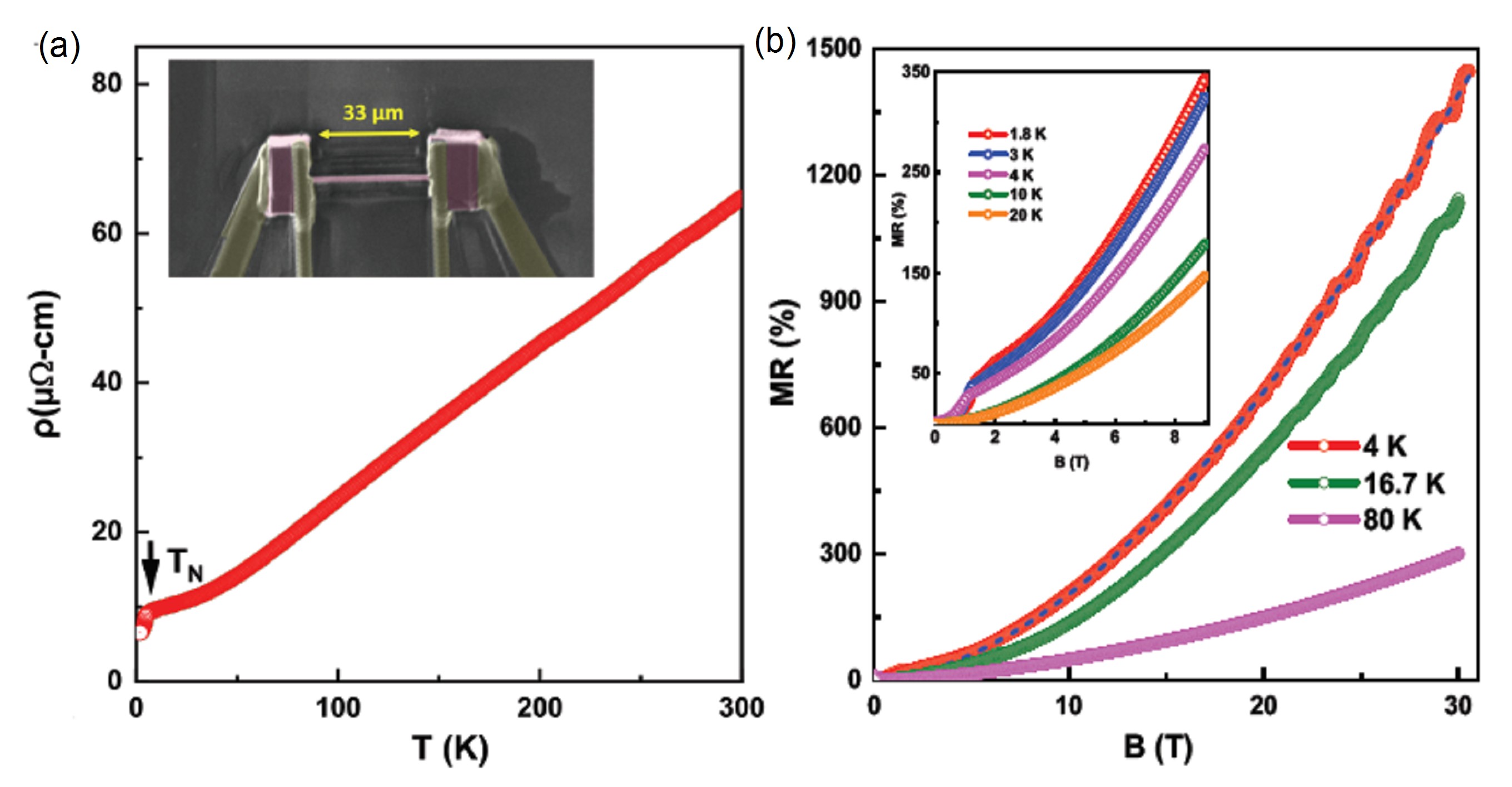}
	\caption{\textbf{Large and non-saturating magnetoresistance in HoSb.} (a) Temperature dependence of electrical resistivity of a PFIB-prepared micrometer-sized single crystal of HoSb. The Inset shows a picture of the PFIB prepared sample (33 $\mu$m $\times$ 0.8 $\mu$m $\times$ 10 $\mu$m). (b) The magnetic field dependence of magnetoresistance of HoSb for applied fields up to 30 T. The dashed line represents the relation MR$\sim$B$^{1.75}$. Inset shows the data in the low-field region. }  
\end{figure*}
\noindent
\textbf{Transport signature of electron-hole compensation}\\
Our ARPES measurements indicate that HoSb is a trivial semimetal. In order to study the transport properties of HoSb, we used a micron-sized sample prepared by plasma focused ion beam (PFIB) microscope. Magneto-transport measurements were performed on a FIB$'$ed single crystal, where the electrical current (\textit{i}) was applied along the [100] crystallographic direction and the magnetic field was applied perpendicular to \textit{i} and [100] (see inset in Figure 4(a)). Temperature dependence of electrical resistivity of the HoSb single crystal is shown in Fig. 4(a). The overall behavior of $\rho$(T) and the presence of an antiferromagnetic ordering at \textit{T}$_N$ = 5.7 K agrees well with previous studies \cite{Xia1}. Application of magnetic field strongly increases electrical resistivity and $\rho$(T) curve saturates at lower temperatures (the resistivity plateau) \cite{Xia1}. The field dependent resistivity saturation at low temperatures has not been observed in the non-magnetic members of the REM family \cite{Niu}. The origin of such a plateau has been recently attributed to the presence of a nearly perfect electron-hole carrier compensation and the high mobility of the compensated semimetals \cite{Guo1}. Figure 4(b) and its inset show the magnetic field dependence of transverse magnetoresistance of HoSb at various temperatures. Magnetoresistance denotes the change of the electrical resistance under applied magnetic field defined as, MR = [$R$($B$)-$R$(0)]/$R$(0), where $R$($B$) and $R$(0) are resistances with and without the magnetic field, respectively. As seen from the inset of Fig. 4 (b), at temperatures below and above $T_N$, MR increases with the increasing magnetic field. The kinks in the MR($B$) curves mark a metamagnetic transition that is related to the change in the magnetic structure with the increase of field \cite{Magnetic}. At $T$ = 1.8 K, magnetoresistance is large, reaching 350 $\%$ at 9 T without any sign of saturation, in agreement with previous reports \cite{Xia1}. In this connection, we measured MR in pulsed magnetic fields up to 30 T (Fig. 4(b)), and found MR measured at several temperatures below and above the Neel temperature increases with magnetic field. MR reaches a large values of about 1500 $\%$ at 30 T ($T$ = 4 K) which is comparable to the values observed in topological semimetals LaSb \cite{LaSb}, NbP \cite{NbP}, and WTe$_2$ \cite{Mazhar}. The MR($B$) curve at 4 K can be described by the form MR $\propto$ $B^n$ (see dashed line in Fig. 4b), where $n$ = 1.75. The close proximity to quadratic field dependence indicates nearly complete electron-hole compensation in HoSb, as expected from a semiclassical two-band model \cite{Guo1}. In addition, for $T$ = 4 K and 16 K, obvious Shubnikov-de Haas (SdH) oscillations are observed at high fields. The presence of the quantum oscillations, even at 16 K, points to the high quality of our HoSb single crystals used in the present studies. Analysis of SdH data in fields up to 30 T will be shown elsewhere. It is also the characteristic of low effective mass and large mobility of carriers that could exist in HoSb. The large, non-saturating magnetoresistance, the resistivity saturation in the turn-on temperature behavior at very low temperatures together with metallic conductivity observed in HoSb are all comparable to the characteristic of topological electron-hole-compensated semimetals \cite{Guo1, ehcom}.\\~\\
\textbf{Discussion}\\
We have carried out in-depth photoemission and transport measurements on HoSb single crystals along with parallel first-principles modeling of the electronic structure of this member of the REM family. Our ARPES measurements show the presence of Dirac-like cone with a small gap at around 470 meV below the chemical potential, but our analysis shows that HoSb is a trivial and not a topological Dirac semimetal. The observed experimental bulk band gap at X point is around 120 meV which is consistent with the theoretically predicted value. Magnetoresistance is found to be large and non-saturating, even at a magnetic field as large as 30 T. The characteristic behavior of electrical resistivity at low temperatures indicates that HoSb is likely a new electron-hole-compensated semimetal.  Notably, the XMR effect has also been reported in other nontrivial members of the REM family suggesting that the presence of non-trivial state might not be directly responsible for the XMR. However, the high carrier mobility associated with the topological states might play a role in generating the XMR. {Furthermore, our magnetoresitivity measurements up to 30 T show a similar unsaturated behavior below and above \textit{T$_N$} indicating similarity of the overall Fermi surface topology in the vicinity of Fermi energy above and below the Neel temperature. Keeping in mind that the rare-earth monopnictide family can host complex magnetic structures including the possibility of a Devil's Staircase transition \cite{DS}, HoSb offers a unique platform for exploring the interplay between XMR, magnetism and topology in an antiferromagnetic matrix.}\\~\\
 \textbf{Methods}\\
 \textbf{Crystal growth and characterization}\\
 Single crystals of HoSb were grown by the Sn flux technique as described elsewhere \cite{Growth}. The crystal structure was determined by X-ray diffraction on a Kuma-Diffraction KM4 four-circle diffractometer equipped with a CCD camera using Mo K$\alpha$ radiation, while chemical composition was checked by energy dispersive X-ray analysis using an FEI scanning electron microscope equipped with an EDAX Genesis XM4 spectrometer.\\~\\
 \textbf{Spectroscopic characterization}\\
 Synchrotron-based ARPES measurements of the electronic structure were performed at the Advanced light Source (ALS) beamline 4.0.3 with a Scienta R8000 hemispherical electron analyzer. The samples were cleaved in situ in a ultra high vacuum conditions (5$\times$10$^{-11}$ Torr) at 18 K. The energy resolution was set to be better than 20 meV. The angular resolution was set to be better than 0.2$^\circ$ for the synchrotron measurements.\\ ~\\
 \textbf{Transport measurements}\\
 For transport measurements, the HoSb sample was prepared by a Plasma Focused Ion Beam (PFIB) method. The electrical resistivity and magnetoresistivity were measured using a Quantum Design Dynacool-9 device in magnetic field up to 9 T. Magnetoresistance measurements at pulsed magnetic fields up to 30 T were performed at NHMFL, and LANL using the same HoSb sample as that used in DC fields. For both measurements, a standard four probe measurement technique was applied.\\~\\
 \textbf{Theoretical calculations}\\
 Electronic structure calculations were performed within the framework of the density-functional theory (DFT) using the projector-augmented-wave (PAW) method \cite{PAW} as implemented in the VASP suite of codes \cite{dft, vasp}. The exchange-correlation functional was treated using SCAN meta-GGA \cite{SCAN}. An energy cutoff of 400 eV was used for the plane-wave basis set and a $\Gamma$-centered 11$\times$11$\times$11 k-mesh was used for BZ integrations. In order to examine the topological properties, we constructed a tight-binding model with atom-centered Wannier functions using the VASP2WANNIER90 interface \cite{Wannier}. The surface energy spectrum was obtained by using the iterative Green’s function method via the WannierTools package \cite{DFT,Green2,Green1}.\\~\\
 %The first-principles calculations were performed within the density functional theory framework using the Vienna {\it ab-initio} simulation package (VASP) \cite{dft,vasp}. The generalized-gradient-approximation with Perdew-Burke-Ernzerhof (PBE) parameterization and projector augmented-wave (PAW) \cite{PBE,PAW} pseudopotential were used in the computations. To calculate topological properties, we generated a real-space tight-binding (TB) model using VASP2WANNIER90 interface \cite{Wannier}. The surface energy spectrum and constant energy contours were obtained by using an iterative Green’s function approach \cite{Green1, Green2}.\\~\\
 % The energy cut-off 400 eV and 9 $\times$ 9 $\times$ 9 k mesh were used to calculate the bulk band structure. A real space TB model based on the Wannier function of As \textit{p} orbitals was built by using WANNIER90 \cite{Wannier} package. The TB model and Green's function \cite{Green1, Green2} method were employed to calculate the surface band structure and the Fermi surface energy contours.\\~\\
\noindent
\textbf{Data availability}\\
All data will be made available by the corresponding author upon request.\\

\bigskip
 \textbf{Acknowledgment}\\
 %M.N. and T.D. at LANL acknowledge support from Department of Energy, Office of Basic Energy Sciences, Division of Material Sciences, and LANL LDRD program.
 M.N.\ is supported by the Air Force Office of Scientific Research under Award No. FA9550-17-1-0415 and the Center for Thermal Energy Transport under Irradiation, an Energy Frontier Research Center funded by the U.S. DOE, Office of Basic Energy Sciences.
 % and LANL LDRD Program.
 %T.D. was supported by Department of Energy, Office of Basic Energy Sciences, Division of Materials Sciences, and by NSF IR/D program.
 D.K., O.P. and P.W. are supported by the National Science Centre (Poland) under research grant 2015/18/A/ST3/00057. O.P. is supported by the Foundation for Polish Science (FNP), program START 66.2020.
 %P.M., A.A., and P.M.O. acknowledge support from the Swedish Research Council (VR), the R\"ontgen-{\AA}ngstr\"om Cluster, and the Swedish National Infrastructure for Computing (SNIC).
 %I.B. acknowledges the support of the NSF GRFP.
 %Work at Princeton University is supported by the Emergent Phenomena in Quantum Systems Initiative of the Gordon and Betty Moore Foundation under Grant No.\ GBMF4547 (M.Z.H.) and by the National Science Foundation, Division of Materials Research, under Grants No.\ NSF-DMR-1507585 and No.\ NSF-DMR-1006492.
 The work at Northeastern University was supported by the US Department of Energy (DOE), Office of Science, Basic Energy Sciences grant number DE-FG02-07ER46352, and benefited from Northeastern University's Advanced Scientific Computation Center (ASCC) and the NERSC supercomputing center through DOE grant number DE-AC02-05CH11231. K.G. acknowledges support from the DOE's Early Career Research Program. N.P acknowledges support from INL's LDRD program (18P37-008FP). Work performed at the NHMFL is supported by NSF cooperative agreement No. DMR-1644779, the State of Florida and DOE. This research used resources of the Advanced Light Source, a U.S. Department of Energy Office of Science User Facility, under Contract No. DE-AC02-05CH11231.
 We thank Sung-Kwan Mo and Jonathan Denlinger for beamline assistance at the LBNL. \\~\\
 \textbf{Author contributions}\\
 M.N. conceived the study; D.K., P.W. and O.P. synthesized the single crystals and performed the transport characterizations; K.G., N.P., F.W. and D. M. prepared the Plasma Focused Ion Beam sample and performed magnetotransport measurements; M.M.H. and M.N. performed the measurements and analysis with the help of G.D., K.D., F.K., C.S., S.R., and W. N.; M.M.H. and M.N. performed the figure planning; B.W., B.S., A.B.S., A.A, and A.B. performed and analyzed the first-principles calculations; A.B. was responsible for the theoretical research direction; M.M.H. and M.N. wrote the manuscript with input from all authors; M.N. was responsible for the overall research direction, planning and integration among different research units.\\~\\
 \textbf{Competing interests}\\
 The Authors declare no Competing Financial or Non-Financial interests.\\~\\
Correspondence and requests for materials should be addressed to M.N. (Email: Madhab.Neupane@ucf.edu).
\newpage

\setcounter{figure}{0}

\renewcommand{\figurename}{\textbf{Supplementary figure}}

\clearpage
\textbf{
	\begin{center}
	\underline{Supplementary Information}\\Observation of gapped state in rare-earth monopnictide HoSb
	\end{center}
	}
	
	\textbf{This file includes}\\
	\textbf{	Supplementary figure 1: Fermi surface and constant energy contour plots.\\
		Supplementary figure 2: Calculated Fermi surface and constant energy contour plots of HoSb.\\
	    Supplementary figure 3: Dispersion maps along the $\Gamma$-M-$\Gamma$ high symmetry direction.\\
		Supplementary figure 4: Observation of bulk bands around the zone center.\\
		Supplementary figure 5: Experimental photon energy dependent dispersion maps along the X-M-X direction.\\
		Supplementary figure 6: Confirmation of gapped state.\\
		%	Supplementary figure 7: Quantitative analysis of Rashba-state.\\
		Supplementary note 1: Fermi surface and constant energy contour plots of HoSb\\
		Supplementary note 2: Observation of the gapped state\\
		%	Supplementary note 3: Quantitative evidence of surface Rashba-state\\
		%	Supplementary note 4: Origin of Rashba spin split surface state in HoSb\\
}

	\clearpage
	\noindent	
	\begin{figure*}
		\centering
		\includegraphics[width=18.5cm]{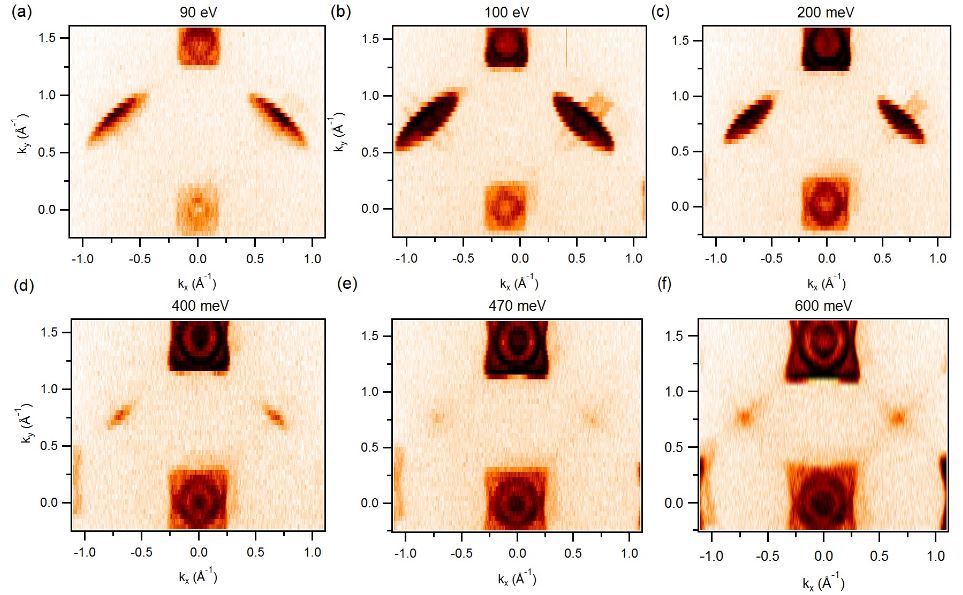}
		\caption{\textbf{Fermi surface and constant-energy contour plots.}	(a),(b) Fermi surface maps of HoSb using various  photon energies (noted in the plots) on a different batch of samples than those discussed in the main text. Photon energies are noted in the plots. (c)-(f) Constant energy contour plots at various binding energies for 100 eV photon energy. Experiments were performed at the ALS beamline 4.0.3 at a temperature of 18 K.
		}
	\end{figure*}

	\noindent
	\begin{figure*}
		\centering
		\includegraphics[width=18.5cm]{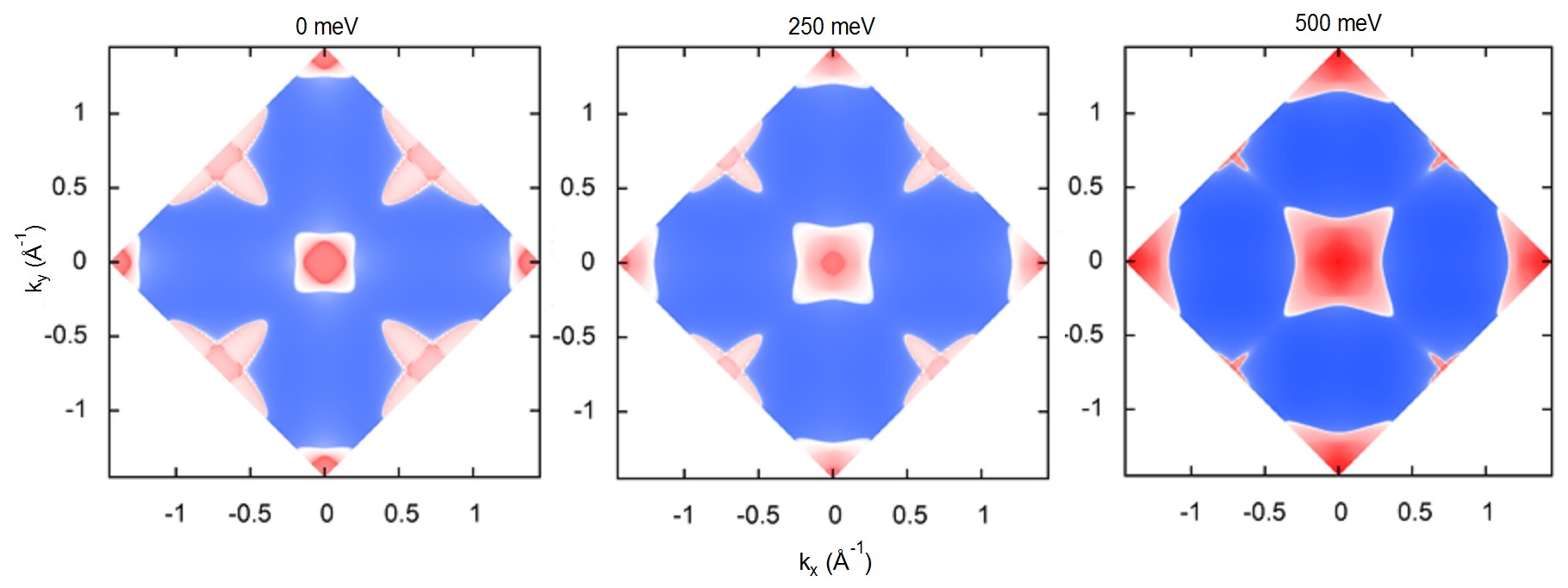}
		\caption{\textbf{Calculated Fermi surface and constant energy contour plots of HoSb.}	 Calculated Fermi surface and constant energy contour plots comparison of HoSb for various values of the binding energy (marked on the plots).
		}
	\end{figure*} 

	\noindent
	\begin{figure*}
		\centering
		\includegraphics[width=18cm]{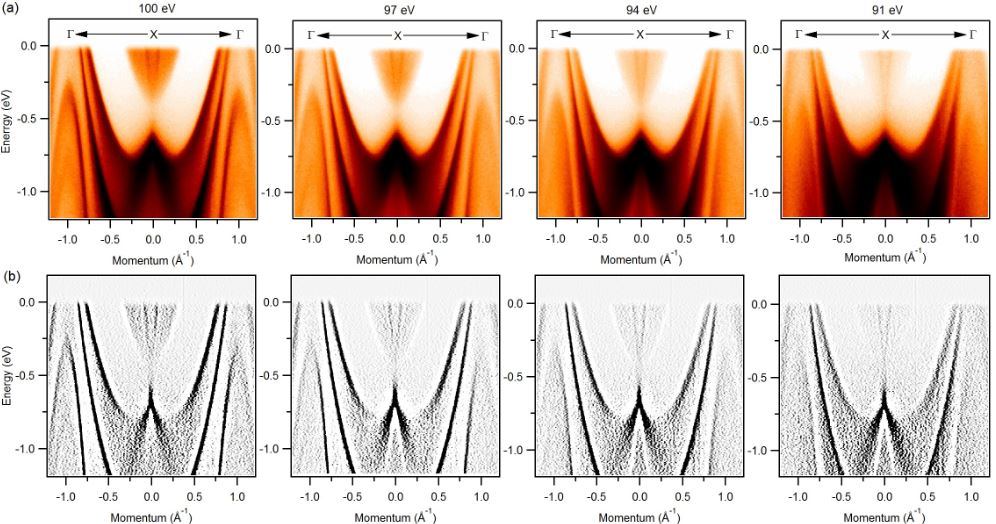}
		\caption{\textbf{Dispersion maps along the $\Gamma$-M-$\Gamma$ high symmetry direction.} (a),(b) Measured dispersion maps along the $\Gamma$-X-$\Gamma$ direction at various photon energies. (b) Second derivative plots for the spectra in (a) obtained by using the curvature methods . Photon energy values are noted in the plots. Experiments were performed at the ALS end-station 4.0.3 at a temperature of 18 K.}
	\end{figure*} 
	\newpage
	\noindent
	\begin{figure*}
		\centering
		\includegraphics[width=18.5cm]{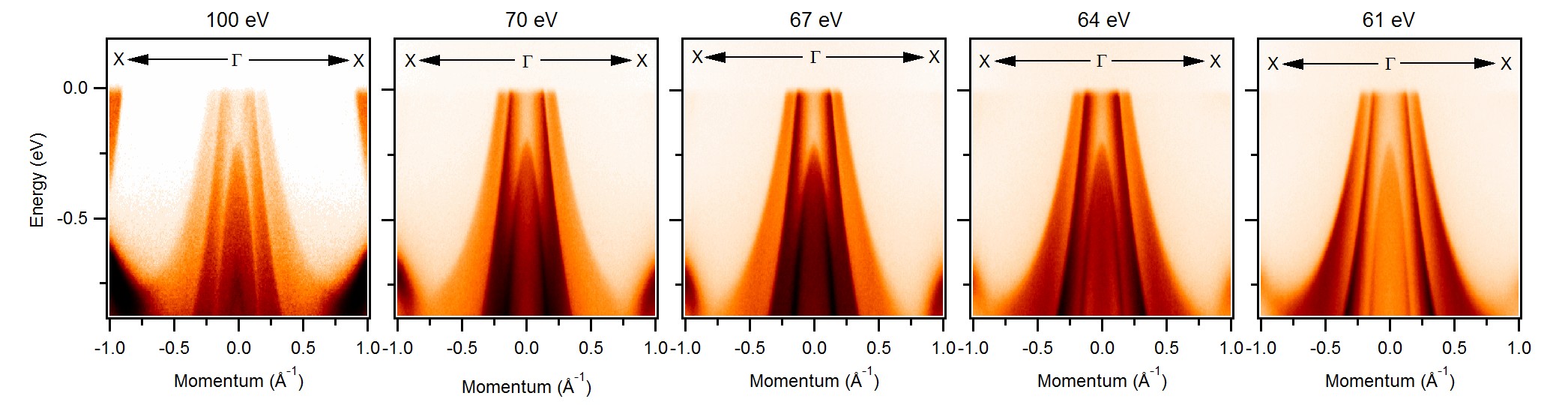}
		\caption{\textbf{Observation of bulk bands around the zone center.} Photon energy dependent dispersion maps along the X-$\Gamma$-X direction. Bands around the zone center ($\Gamma$) show notable dispersion as a function of incident photon energy. Measurements were performed at the ALS beamline 4.0.3 at a temperature of 18 K.}
	\end{figure*} 
	\noindent
	
	\noindent
	\begin{figure*}
		\centering
		\includegraphics[width=18cm]{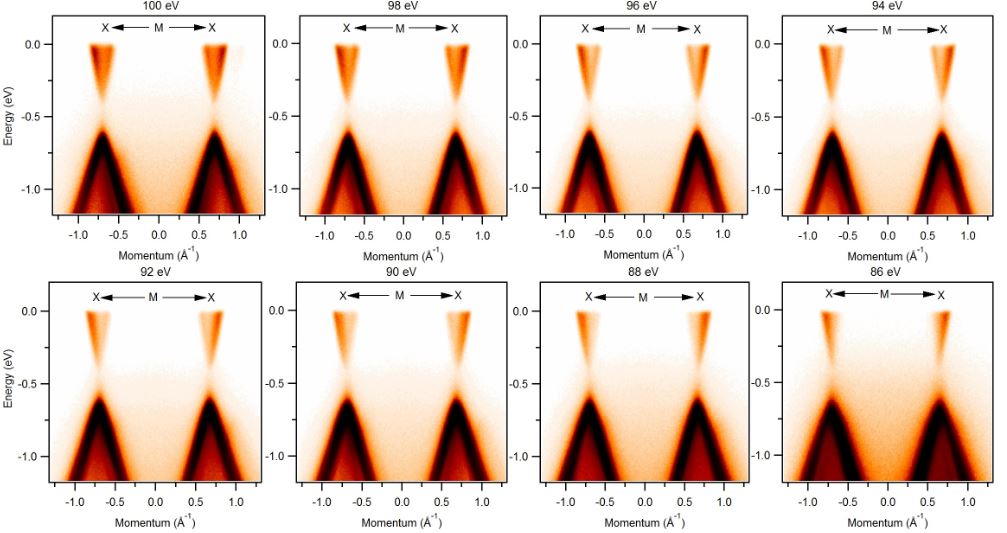}
		\caption{\textbf{ Experimental photon energy dependent dispersion maps along the X-M-X direction.} Measured dispersion maps along the X-M-X direction using various photon energy with 2 eV energy steps from 100 eV to 86 eV as noted over the plots. Measurements were performed at the ALS beamline 4.0.3 at a temperature of 18 K. }
	\end{figure*}
	\noindent
	\begin{figure*}
		\centering
		\includegraphics[width=16cm]{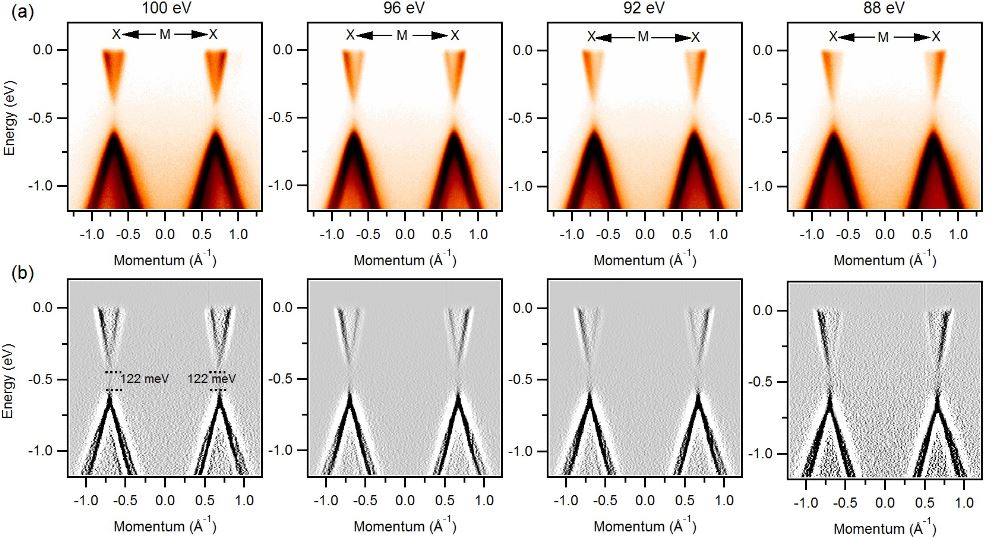}
		\caption{{\textbf {Confirmation of gapped state}. }(a),(b) ARPES measured dispersion maps along the high symmetry X-M-X direction and second derivative plots with different photon energies, respectively. Gapped state can be clearly seen. All measurements were performed at the ALS beamline 4.0.3 at a temperature of 18 K.}
	\end{figure*} 
	\newpage
	
	\noindent
	\textbf{Supplementary Note 1}\\
	\textbf{Fermi surface and constant energy contour plots of HoSb}\\
	We measured a second batch of samples to access the robustness of our Fermi surface and electronic structures results. Supplementary Figure (SF) 1 shows the observed Fermi surface (SF. 1(a)-(b)) and the related constant energy contour plots (SF. 1(c)-(f)) for the new measurements. Like the Fig. 1 in the main text, Fermi surface consists of an outer diamond and an inner circular pocket at the $\Gamma$ point (zone center) and two concentric elliptical pockets at the X point of the Brillouin zone (BZ) (SF. 1(a)-(b)). Importantly, at a higher binding energy ($\sim$470 meV), we observe that the elliptical-pocket evolves into a point-like feature. Supplementary Figure 2 shows the corresponding calculated Fermi surface and the related constant energy contour plots. An excellent agreement is seen between the experimental data and theoretical predictions.\\~\\
	\textbf{Supplementary Note 2\\ Observation of the gapped state}\\
	In order to determine the origin of the bands near the zone center and the corner of the BZ, we performed photon energy dependent dispersion maps around these high-symmetry points. Supplementary Figures 3(a) and,(b) show the measured photon energy dependent dispersion maps and their second derivative plots along the $\Gamma$-X-$\Gamma$ direction, respectively. Around the $\Gamma$ point, two hole-like bands cross the Fermi level while at the X point, we observe a nearly linearly dispersive feature. To confirm the origin of the bands near the $\Gamma$ point, we present more photon energy dependent dispersion maps in Supplementary Fig. 4 over a wide energy window. Here, one can clearly observe the photon energy dependent dispersive nature of the bands, so that the hole-like bands at around the zone center are bulk originated. Furthermore, from Supplementary Figure 3(b), we see a clear gap at the X point. In order to further confirm the gapped state at the X point, we present photon energy dependent dispersion maps along the X-M-X direction in the supplementary Fig. 5. The gapped state is seen consistently at all photon energies. In order to observe the gap size more vividly, we present measured dispersion maps and their second derivative plots in Supplementary Fig. 6. From panel 6(b), we approximate the band gap size around 120 meV. Moreover, the bulk state is once again clearly observed to be gapped which is consistent with the main text results.\\~\\
\end{document}